\documentclass[doublecol]{epl2}
\usepackage{ulem}

\def\be{\begin{equation}}
\def\ee{\end{equation}}     
\def\bfi{\begin{figure}}
\def\efi{\end{figure}}
\def\bea{\begin{eqnarray}}
\def\eea{\end{eqnarray}}
\newcommand{\comment}[1]{}

\begin{document}

\title{Effective mobility and diffusivity in 
coarsening processes}

\author{Federico Corberi\inst{1} \and Eugenio Lippiello\inst{2} \and Paolo Politi\inst{3,4}}
\institute{
\inst{1}Dipartimento di Fisica ``E.~R. Caianiello'', and INFN, 
Gruppo Collegato di Salerno, and CNISM, Unit\`a di Salerno,Universit\`a  di Salerno, 
via Giovanni Paolo II 132, 84084 Fisciano (SA), Italy.\\
\inst{2} Dipartimento di Matematica e Fisica, Universit\`a della Campania "L. Vanvitelli", Viale Lincoln, Caserta, Italy.\\
\inst{3} Istituto dei Sistemi Complessi, Consiglio Nazionale
delle Ricerche, via Madonna del Piano 10, I-50019 Sesto Fiorentino, Italy.\\
\inst{4} Istituto Nazionale di Fisica Nucleare, Sezione di Firenze, via G. Sansone 1 I-50019, Sesto Fiorentino, Italy.
}


\abstract{
We suggest that  coarsening dynamics can be  described in terms of a generalized random walk, with the dynamics of  the growing length $L(t)$ controlled by a drift term, $\mu(L)$,  and a diffusive one, ${\cal D}(L)$.
We apply this interpretation to the one dimensional Ising model 
with a ferromagnetic coupling constant decreasing exponentially on the scale $R$.
In the case of non conserved (Glauber) dynamics, both terms are present and their
balance depends on the interplay between $L(t)$ and $R$. In the case of conserved (Kawasaki) dynamics,
drift is negligible, but ${\cal D}(L)$ is strongly dependent on $L$.
The main pre-asymptotic regime displays a speeding of coarsening
for Glauber dynamics and a slowdown for Kawasaki dynamics.
We reason that a similar behaviour can be found in two dimensions.
}

\maketitle

\section{Introduction}
Driven systems and systems relaxing towards equilibrium may display
a common dynamical feature, coarsening~\cite{crp}, namely the increase in time of a typical length, $L(t)$,
which represents some peculiar property of the system under study.
The most classic example is phase ordering following a temperature
quench across a critical point \cite{quench}: in this case dynamics is fully dissipative and it is driven by energy minimization. 
Other standard examples concern pattern forming systems \cite{pattern}, where coarsening follows a phase
instability of periodic structures, condensation phenomena \cite{condensation}, and 
nonequilibrium phase separation processes \cite{quench}.
Coarsening is also seen to accompany energy localization in systems with conservation laws,
through the maximization of entropy~\cite{dnls}. 

A simple and fundamental description of coarsening in binary systems is provided by the
ferromagnetic Ising model with a generic interaction constant $J(r)>0$ between two spins
at distance $r$. In particular, we will focus on a coupling
$J(r) \sim \exp(-r/R)$, where $R$ is an interaction range.
Coarsening is observed when the system is quenched from an initially disordered
configuration corresponding to a high temperature equilibrium state, to a temperature
$T$ well below the critical temperature (or to $T\simeq 0$ in one dimension).
Depending on the system at hand, the dynamics can conserve
the order parameter or not. The former case (COP$=$Conserved Order Parameter) 
amounts to a lattice gas model,
where only spin exchanges are possible, which provides a correct description
of binary mixtures. The latter case (NCOP$=$Non Conserved Order Parameter)
is adequate to mimic
magnetic materials, where the evolution proceeds by single spin flips.
A representation of these two models is provided in Fig.~\ref{fig1}
where some simple configurations in dimension $d=1$, to be discussed later, are drawn.

The goal of this Letter is to relate the behavior of $L(t)$
to a biased diffusion process of a generalized Brownian motion, through the inverse relation
\be
\frac{1}{t(L)} =\frac{\mu(L)}{L} + \frac{2{\cal D}(L)}{L^2},
\label{etL}
\ee
where $t(L)$ is the average time needed to a fictitious walker to move a distance $L$. 
Our approach provides a rather simple effective description of 
the complex phase-ordering phenomenon, whereby many degrees of freedom strongly 
interact in a non-linear way, in terms of a single diffusing variable described
by Eq. (\ref{etL}).  The asymmetric  and symmetric parts of the Brownian motion
contribute to the first and second term in the above equation, respectively.
It is straightforward that 
the standard ballistic motion corresponds
to a constant $\mu$ and a vanishing ${\cal D}$,
while the diffusive motion is obtained in the opposite case,
constant ${\cal D}$ and vanishing $\mu$.

The Ising model with nearest neighbourgh interaction
is known to display the coarsening laws
$L(t)\propto t^{1/2}$ (NCOP) and $L(t)\propto t^{1/3}$ (COP).
Such relations are due to a vanishing $\mu(L)$ and to 
${\cal D}(L)\simeq 1$ (NCOP) or ${\cal D}(L)\simeq 1/L$ (COP).
These behaviors for the drift and the diffusivity
will be recovered asymptotically ($t\to\infty$) also for $J(r) \sim\exp (-r/R)$,
but for large $R$ the system displays a variety of different
dynamical regimes.
Our calculations allow us to determine analitically
the form of the growth law $L(t)$ 
for the one dimensional model along the whole history of the system,
from the quench instant up to the asymptotic stages, thus revealing
the existence and the features of such regimes. 
Furthermore, in the framework provided by Eq.~(\ref{etL}) the interpretation of the complex 
evolution occurring in the coarsening system is made more transparent,
allowing one to appraise the role of the conservation law.

For example, we will show that $\mu$ plays a prominent role in expediting
coarsening preasymptotically with NCOP, while the absence of any drift 
together with a vanishing ${\cal D}$ makes the
COP early dynamics sluggish at large $R$. 
Moreover, although the existence of a slow logarithmic 
coarsening, $L(t) \sim \ln t$, will be shown to occur both with COP and NCOP, the interpretation
of these two apparently similar behaviors in the framework of Eq.~(\ref{etL}) unveils
a fundamental physical difference: in the former case it is due to
an unbiased diffusion with an exponentially small diffusivity, while in the latter case 
it is caused by an exponentially
small drift.
In this Letter, besides developing the analytical arguments described insofar,
we compare them with the outcome of numerical simulations, finding agreement.
In two dimensions we are able to support a similar picture, either with numerics or
with physical considerations.

\bfi
\begin{center}
\includegraphics*[width=0.95\columnwidth,clip=yes]{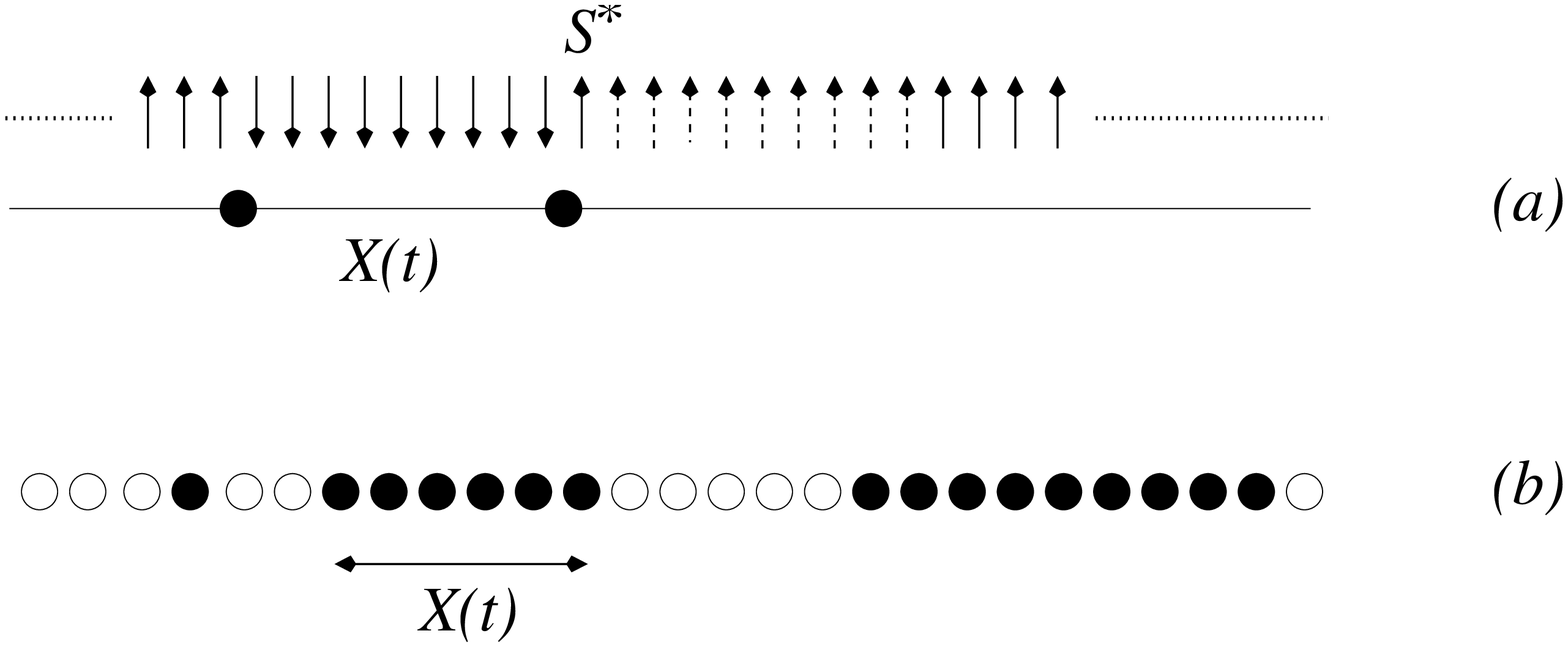}
\end{center}
\caption{
Simple one dimensional configurations. 
(a) A single domain of down spins of length $X$ within a sea of up spins.
The interaction of spin $S^*$ with down spins is compensated by its
interaction with dashed up spins.
(b) Two clusters of particles, whose size changes in time because they exchange monomers
(periodic boundary conditions apply).
}
\label{fig1}
\efi

\section{General results from simple one dimensional configurations}

In the following we will use the simple configurations of Fig.~\ref{fig1}
to derive the coarsening laws of a system in an initially disordered
state. Although the latter contains many domains and is much more
complex than the ones of Fig.~\ref{fig1},
the study of simplified configurations
is justified by the existence of dynamical scaling.  According to that
at a given time $t$, the system is characterized by a single length scale $L(t)$,
whose time dependence is the so-called coarsening law. Due to that,
instead of evaluating the average size $L$ of domains at time $t$ in the actual
configuration of the system, we can reverse the
line of reasoning, and evaluate the
time $t$ required for the annihilation of a domain
of size $L$ in the simplified configurations of Fig.~\ref{fig1}.
In both cases, NCOP and COP, we have a quantity $X(t)$,
either the distance between two neighbouring domain walls or the size of a cluster,
which performs a random walk and in both cases coarsening occurs because 
eventually $X(t)$ vanishes, thus decreasing the domain wall or cluster density.

We start from the
case with NCOP.
In Fig.~\ref{fig1}(a)
a single domain of down spins is immersed in a sea of up spins. 
The evolution proceeds by spin flips,
and can be more easily formulated in terms of motion of domain walls, black full dots in the figure.
At zero or very low temperature the only allowed processes correspond to the hopping of interfaces
and to their annihilation, while the creation of new domain walls is energetically expensive.
In particular, coarsening occurs because interfaces annihilate upon meeting.

Let us consider the spin $S^*$, see Fig.~\ref{fig1}(a).
Its interaction energy with the cluster of down spins to its left is offset by the interaction energy with
(dashed) up spins within a distance $X$ to its right. Since all remaining spins are up, flipping 
$S^*$ costs some energy while  reversing the spin to 
its left lowers the energy. In conclusion, the domain wall jumps asymmetrically, 
with an $X$-dependent drift that favours decreasing $X$, thus speeding up coarsening.
If the interaction is limited to nearest neighbour this mechanism is not active.
In general, the drift is locally effective if domains are smaller than the interaction range.

Using a convection-diffusion equation  to describe the evolution of $X(t)$, if $X(0)=L$, we find 
\cite{redner} that the time $t$ required to close
the domain, $X(t)=0$, is given by
\be
t(L) = \frac{L}{v} \tanh\left( \frac{vL}{2D} \right),
\label{tL}
\ee
where the constant $D$ is the diffusion coefficient of the domain wall, which
can be assumed to be constant, and $v$ is its drift.
The drift $v(L)$ is proportional to the asymmetry of the hopping,
$v(L)=v_0\delta(L)$, where $v_0$ is a constant.
If $p_\pm$ are the probabilities for the domain wall to hop to the right and to the left, respectively,
assuming detailed balance the asymmetry $\delta(L)=p_--p_+$ reads
\be
\delta(L) =\tanh\left(\frac{\beta\Delta E(L)}{2}\right),
\label{delta}
\ee
where
\be
\Delta E(L) = 4\int_L^\infty J(r) dr
\label{deltaEncop}
\ee
is the energy difference due to the spin flip. 
Equations~(\ref{tL},\ref{delta},\ref{deltaEncop}) 
define the closure time $t(L)$ in terms of the 
{\it bare} quantities, the diffusivity $D$ and the drift $v(L)$, which regulate the evolution of $X(t)$.
Instead, $\mu (L)$ and ${\cal D}(L)$ appearing in Eq. (\ref{etL}) are an effective drift 
and diffusivity for $L(t)$, which are different, in principle, from $D$ and $v(L)$.
In order to arrive at explicit expressions for the effective quantities we identify 
$\mu(L)$ with $v(L)$, since this quantity is obviously related to the asymmetry $\delta$. 
Therefore
\be
\left \{
\begin{array}{l}
\mu(L)=v_0\tanh\left(\frac{\beta\Delta E(L)}{2}\right) \\
   {\cal D}(L) = \frac{1}{2}L\mu(L)\left[ \tanh^{-1}\left(\frac{\mu(L)L}{2D}\right) -1\right] .
   \end{array} \right .
\label{summaryncop}
\ee
Once the form of the coupling $J(r)$, hence of $\Delta E$, is specified,
Eqs. (\ref{deltaEncop},\ref{summaryncop}) are a close
set of equations from which $\mu$ and ${\cal D}$ can be extracted.
This, in turn, allows one to determine the growth law $L(t)$ by reversing
Eq.~(\ref{etL}) or Eq.~(\ref{tL}).

Before pursuing this program, let us discuss the
modifications to the present approach needed for a system with COP.
In order to do this let us refer to the configuration of Fig.~\ref{fig1}(b).
In the
lattice gas representation this is interpreted as two clusters of particles
(e.g., up spins),
whose total number is kept constant by the conservation law, separated by empty spaces (down spins).
In this case, dynamics proceeds by detaching of monomers from clusters,
and their further hopping until a neighbouring droplet is met (periodic boundary conditions
are assumed).
This corresponds to a neat exchange
of monomers between particle domains and coarsening occurs because
one of them empties completely in favor of the other.

Let us denote by $X(t)$ the size of a cluster. 
The quantity $X(t)$ fluctuates because of particles exchanges with the neighbouring droplet.
In first approximation the exchange of particles between neighboring clusters is a
symmetric process, which implies $\mu (L)=0$.
If each monomer detached from a domain has a finite, constant probability 
$2D$ to attach to the neighbouring one, $X(t)$ would simply perform a standard random walk, 
with a diffusivity $D$,
and the time $t(L)$ to close a cluster of initial size $X(0)=L$ would be equal to 
$L^2/2D$. Instead, 
the detached monomer has a probability $2{\cal D}(L)$ to attach to the neighbouring cluster which depends on
their distance, because before arriving to the nearest domain it can return back. This probability 
scales with the size $L$ of the cluster itself, giving
$t(L) = L^2/2{\cal D}(L)$.
If $J(r)$ is not restricted to a nearest neigbour coupling, once the particle has detached it feels 
an initial drift
to return back. This fact strongly reduces the probability to attach to the neighbouring cluster,
slowing down coarsening.

If the drift $-v$ felt by the travelling particle were constant, 
the probability $p(x_0,\ell)$ that after being released in $x=x_0$ 
it attains the point $x=\ell$ would be given by
\be
p(x_0,\ell) = \frac{e^{\frac{v x_0}{D}} -1}{e^{\frac{v\ell}{D}} -1} ,
\ee
where $D$ is the bare diffusivity of the travelling particle.

In reality, the drift depends on the distance between the monomer and
the cluster of origin: it is maximal 
after detaching, then decreases, finally changing sign
when the particle attains the midpoint $L/2$ between the two clusters. 
Therefore, once the midpoint is passed, the attachment to the neighbouring cluster
is almost certain. For this reason 
we can evaluate ${\cal D}(L)$ through the relation $2{\cal D}(L)=p(1,L/2)=
\frac{e^{\frac{v}{D}} -1} {e^{\frac{vL}{2D}} -1}$.

The drift $v$ is again proportional to the asymmetry $\delta (L)$ defined
above Eq. (\ref{deltaEncop}), where now 
$\Delta E(x,L)$ is the variation of energy $E(x)$ of the monomer when it moves
from position $x$ to $x+1$. 
In a continuum picture $E(x) = \int_0^\infty dy [J(x+y) + J(L-x+y)]$, where 
we have neglected the interaction on scales larger than $L > R$. Then,
\be
\Delta E(x,L) = -\int_0^\infty dy \,[J'(x+y) -J'(L-x+y)].
\label{Delta_COP}
\ee
where $J'(x)$ is the spatial derivative of the coupling costant $J(x)$. 
Notice that, at variance with NCOP, the asymmetry $\delta$ now depends also on $x$. 
We simplify the problem by estimating the drift using the spatial average $\delta (x)$, hence
\be
v=v_0\frac{2}{L}\int _0^{L/2}\delta (x) \, dx ,
\label{copv}
\ee
and we arrive at
\be
\left \{
\begin{array}{l}
\mu(L)=0 \\
{\cal D}(L) = \frac{1}{2}\,\frac{e^{\frac{v}{D}} -1} {e^{\frac{vL}{2D}} -1}.
   \end{array} \right .
\label{summarycop}
\ee
Equations~(\ref{copv},\ref{summarycop}), together with the definition (\ref{delta})
and expression (\ref{Delta_COP}) are the close set of equations for COP,
analogue to Eqs. (\ref{delta}-\ref{summaryncop}) for NCOP,
allowing one to infer the growth law once an explicit form 
for $\Delta E$ is given. 
In this respect Eqs. (\ref{delta}-\ref{summaryncop})
and Eqs. (\ref{copv},\ref{summarycop}) are quite general.

\section{Growth laws}

Let us now show how the considerations of the previous section can be used
to predict the growth law of $L(t)$ in the whole one dimensional
system with many interfaces.
Specifically, we consider the model with
Hamiltonian
\be
{\cal H}=-\sum _{(ij)}J_{ij}S_iS_j
\ee
where $S_i=\pm 1$ are Ising spins on a regular lattice.
The indexes $i$ and $j$ in the sum run over all the distinct pairs $(ij)$ of lattice sites
and the coupling constant is chosen as~\cite{exp_sys}
\be
J_{ij}=J(r)=J_0e^{-r/R},
\label{coupling}
\ee
where $r=\vert i-j \vert$ is the distance between two sites,
$R$ is an interaction range, and $J_0$ will be taken in the following equal to one.
The coarsening kinetics is induced by preparing the system in a fully disordered
equilibrium state at $T_i=\infty$ and quenching it at time $t=0$ to a
sufficiently low final temperature $T_f$.
The system evolves by flipping single spins (Glauber dynamics) or couples
of neighboring antiparallel spins (Kawasaki dynamics), in the case of
NCOP or COP, respectively.

We have studied numerically the system described above, using Glauber transition rates
for spin flips. We start presenting our data in $d=1$.
The system size in this case is $N=10^7$. This value is sufficiently
large to avoid any finite-size effect in the range of simulated times.
We compute the average size $L(t)$ of the growing domains as the inverse
density of antialigned nearest neighbor spins~\cite{quench} (the same method will be used in $d=2$
further on).

\subsection{NCOP}

\begin{figure}[h]
\begin{center}
\includegraphics*[width=0.95\columnwidth]{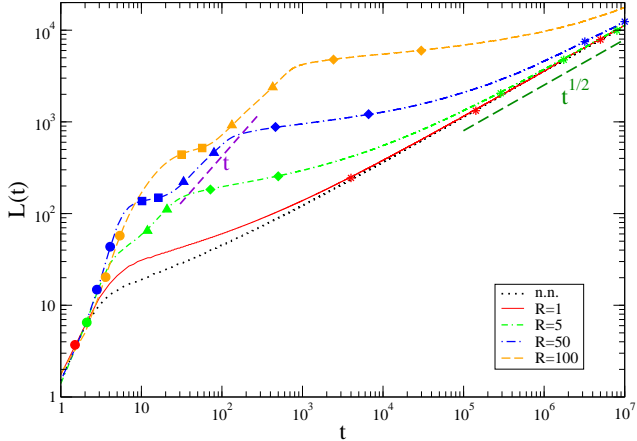}
\end{center}
\caption{The domain size $L(t)$ is plotted against time on a double logarithmic
  scale for a quench with NCOP from $T_i=\infty$ to $T_f=10^{-2}$, in $d=1$.
  Different curves correspond to the next neighbors (n.n.) interaction and to the 
  coupling of Eq.~(\ref{coupling}) with different values of the interaction range $R$ (see the legend).
  Different symbols indicate different dynamical regimes, see text.
  The dashed green and violet lines are the algebraic forms $t^{1/2}$ and $t$,
respectively.} 
\label{fig_onekinkreg}
\end{figure}

The evolution of $L(t)$ after quenching a system with
NCOP to $T_f=10^{-2}$, 
is shown in Fig. \ref{fig_onekinkreg}.
This figure shows the existence of five temporal regimes where $L(t)$ grows
in a markedly different way. Such regimes are highlighted by using,
for each curve, different symbols. We denote the first two regimes, marked by
circles and squares respectively, as the {\it exponential} and {\it plateau}
stages. In these stages $L(t)$ grows first exponentially fast, and than
saturates to an approximately constant value.
Initially $L(t)$ is so small that, upon moving a distance of order $R$,
one encounters many interfaces. 
Therefore, the analytic approach of the previous section, based on
simplified configurations with one or two kinks alone, cannot be applied.
For this reason, we postpone the discussion of this early time behavior to the
last part of the paper. When this initial stage is over, the system enters
a {\it ballistic} (marked with triangles), then a {\it logarithmic} (diamonds), and eventually a {\it diffusive}
regime (stars), where $L(t)$ increases linearly, logarithmically,
and as $t^{1/2}$, respectively. Let us see now how the growth law $L(t)$ can
be predicted in these regimes.

{\it Ballistic regime ---} For sufficiently large values of their arguments, the two hyperbolic tangents
appearing in Eqs. (\ref{summaryncop}) attain a unitary value. 
This occurs for $\frac{2D}{v_0}\ll L(t) \ll R\ln (2\beta R J_0)$, where we used Eq.~(\ref{coupling}).
Then $\mu (L)=v_0$ and ${\cal D}(L)\simeq 0$ and, after Eq. (\ref{etL}), $L(t)\simeq v_0t$
grows linearly in time, as it is observed in Fig. \ref{fig_onekinkreg}.

{\it Logarithmic regime ---} For sufficiently small values of the argument of the hyperbolic function in the
first of Eqs. (\ref{summaryncop}) one has $\mu(L)\simeq 2v_0\beta RJ_0e^{-L/R}$, where we have used
Eqs. (\ref{deltaEncop},\ref{coupling}). If, at the same time, the argument of the hyperbolic tangent
in the second of Eqs. (\ref{summaryncop}) is sufficiently large one still has ${\cal D}(L)\simeq 0$.
The two conditions expressed above amount to 
$R\ln (2\beta RJ)\ll L(t)\ll R\ln \left [\beta  Jv_0D^{-1}R^2\ln (2\beta JR)\right ]$.
In this range Eq. (\ref{etL}) predicts a logarithmic growth, as observed in Fig. \ref{fig_onekinkreg}.
For large $\beta R$, it reads
$L(t) \simeq (a+\ln t)R$, where $a=\ln(2v_0\beta RJ_0)$.

{\it Diffusive regime ---} For $L(t)\gg R\ln \left [\beta  Jv_0D^{-1}R^2\ln (2\beta JR)\right ]$, from 
Eqs. (\ref{summaryncop}) one finds that $\mu (L)$ is negligible and, using the smallness of
the hyperbolic tangent argument, ${\cal D}(L)\simeq D$. This gives the usual asymptotic 
growth law $L(t)=\sqrt{2Dt}$ of phase-ordering with NCOP, as shown in Fig. \ref{fig_onekinkreg}. 
A summary of the behaviors of $\mu(L)$ and ${\cal D}(L)$, and of the associated growth laws,
is given in Table \ref{table_summary}.  

\subsection{COP}

\begin{figure}[h]
\includegraphics*[width=0.95\columnwidth]{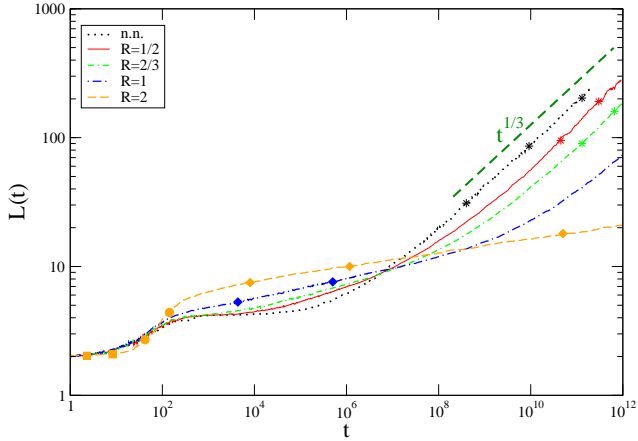}
\caption{$L(t)$ is plotted against time on a double logarithmic
  scale for a quench with COP to $T_f=0.4$ in $d=1$.
  Different curves correspond to the next neighbors (n.n.) interaction and to the 
  coupling of Eq.~(\ref{coupling}) with different values of $R$ (see legend).
  Different symbols indicate different dynamical regimes, see text.
  The dashed green line is the algebraic form $t^{1/3}$.} 
\label{lt_COP_beta2.5}
\end{figure}

The evolution of $L(t)$ after quenching a system with
COP from $T_i=\infty$ to $T_f=0.4$ is shown in Fig. \ref{lt_COP_beta2.5}.
This value of $T_f$ was chosen as a compromise between the wish to reach low temperatures---%
where the various dynamical regimes are clearly observed---and the need to avoid the exponentially slow activated
kinetics occurring at very low $T_f$.
In this case one can appreciate the existence of four different regimes, visible upon tuning $R$:
an initial {\it plateau} regime (marked by squares) where $L(t)$ stays constant, followed by an
{\it exponential} increase of $L(t)$ (circles), a {\it logarithmic} regime (diamonds) and, eventually,
a diffusive stage (stars) where the usual asymptotic growth law $L(t)\sim t^{1/3}$ of conserved
systems is observed. At variance with NCOP there is no ballistic regime and the plateau
occurs before the exponential growth. As for NCOP the first two regimes cannot be 
interpreted with the analytic arguments  of the previous section, which hold for a single domain. 
They will be discussed later.
Let us now discuss the other two regimes.

{\it Logarithmic regime ---} With the coupling (\ref{coupling}), $\Delta E$ (defined in
Eq.~(\ref{Delta_COP})) reads
\be 
\Delta E(x,L)=J_0  \left [e^{-x/R}-e^{-(L-x)/R}\right ], 
\label{DEcop}
\ee
where $x\le L$.
For short times, when $L$ is much smaller than $R$, we can expand the arguments 
of the exponentials to first order. Using the result in Eq. (\ref{delta}) and expanding 
the hyperbolic function therein one has $v(L)\propto L$. We will show below that
$v(L)$ vanishes for long times. Hence we conclude that $v(L)$ attains a maximum value $v_M$
for a certain value $L_M$ of $L(t)$. Letting $v(L)\simeq v(L_M)\equiv v_M$ in the second of Eqs.~(\ref{summarycop})
one has ${\cal D}(L)\simeq \frac{1}{2}\,\frac{e^{\frac{v_M}{D}} -1} {e^{\frac{v_ML}{2D}} -1}$ in this regime. 
Plugging into Eq. (\ref{etL}) one obtains, for sufficiently large $L(t)$, a logarithmic growth law
$L(t)\simeq b\ln t$, where $b$ is a ($R$ independent) constant. 
 
 {\it Diffusive regime ---} For $L\gg R$ the integral in (\ref{copv}) takes contributions only for
 $0\le x \le R$ where, using Eqs. (\ref{delta},\ref{DEcop}), 
 $\delta (L)$ can be roughly evaluated as $\delta(L) \simeq \tanh(\beta J_0/2)$.
 Using this approximated value in Eq. (\ref{copv}) one has $v=2v_0(R/L)\tanh (\beta J_0/2)$  and
 hence $L(t)\simeq c\,t^{1/3}$ (with $c=2(v_0/D)R\tanh(\beta J_0/2)e^{-(v_0/D)R\tanh(\beta J_0/2)}$).
 This is the usual behavior for systems with short range interactions. 
As for NCOP, a summary of the behaviors of $\mu(L)$ and ${\cal D}(L)$, and of the
associated $L(t)$, is given in table \ref{table_summary}.  

\vspace{.5cm}

\begin{table}
\begin{center}
\begin{tabular}{|c|c|c|c|} 
  \hline
  \multicolumn{4}{|c|}{NCOP}\\
  \hline
  & Ballistic & Logarithmic & Diffusive \\
  \hline
  $\mu(L)$ & $v_0$ & $e^{-L/R}$ & Negligible \\
  \hline
  ${\cal D}(L)$ & Negligible & Negligible & $D$ \\
\hline 
  $L(t)$ & $v_0t$ & $R\ln t$ & $t^{1/2}$ \\
  \hline
  \end{tabular}
  \end{center}

\begin{center}
\begin{tabular}{|c|c|c|} 
  \hline
      \multicolumn{3}{|c|}{COP}\\
  \hline
  & \hspace{.52cm} Logarithmic \hspace{.52cm} & \hspace{.6cm} Diffusive \hspace{.6cm}\\
  \hline
  $\mu(L)$ & Negligible & Negligible \\
  \hline
  ${\cal D}(L)$ & $e^{- L}$ & $R/L$ \\
\hline 
  $L(t)$ & $\ln t$ & $t^{1/3}$ \\
  \hline
\end{tabular}
\end{center}
\caption{Summary of the behaviors of $\mu(L)$, ${\cal D}(L)$, and $L(t)$.}
\label{table_summary}
\end{table}

\section{Early stages}

As already mentioned, our analysis of coarsening based on simple
one dimensional configurations with 
few domain walls is not suited to describe the initial stages when $L(t) \ll R$. This corresponds to the
first two regimes for NCOP and COP as well. In the limit $L(t)/R\to 0$ one has infinitely many
interfaces inside the interaction range, a fact that can be regarded as a mean field situation with an
effective Hamiltonian 
\be
{\cal H}=-J_0 m\sum _iS_i,
\label{Hmeanf}
\ee
where $m=\langle S_i\rangle$ is the magnetization (per spin). 

Focusing on the NCOP case,
the probability $p(t)$ of finding $S_i=+1$ at time $t$ obeys the master equation
$dp(t)/dt=-p(t)W_\uparrow+[1-p(t)]W_\downarrow$,
where $W_\uparrow $ and $W_\downarrow $ are the transition 
rates to flip an up or down spin, respectively.
Using $m=2p(t)-1$ and detailed balance with respect to the Hamiltonian (\ref{Hmeanf}), 
one easily arrives at
$dm(t)/dt={\cal W}\left [-m+\tanh (\beta Jm)\right ]$, where
${\cal W}=W_\uparrow +W_\downarrow $ is the inverse of 
the average number of spin flips per unit time.
For $\beta >\beta_c=1/J_0$,
starting with a small magnetization $m(0)$, one has
\be
\left \{
\begin{array}{ll}
m(t)=m(0)e^{{\cal W}(\beta J_0 -1)t}, & t\ll [{\cal W}(\beta J_0-1)]^{-1} \\
m(t)\simeq m_{eq}, & t\gg [{\cal W}(\beta J_0-1)]^{-1}
\end{array}
\right . 
\label{mmeanf}
\ee
where $m_{eq}=\tanh (\beta J_0m_{eq})$ is the equilibrium value.
This implies that the total size of regions where spins are aligned with $m(t)$, a quantity proportional
to $L(t)$, increases in the same way. This explains the first two regimes, denoted above as 
{\it exponential} and {\it plateau}, observed with NCOP.
Notice that our mean field approach applies independently 
in any portion of the system with a size of order $R$, and $m(0)$ is the initial magnetization therein.
Therefore Eq. (\ref{mmeanf}) does not imply that the up-down symmetry is broken in the whole
system, since positive and negative values of $m(0)$ occur with equal probability in different
regions, and the overall magnetization remains negligible.

For COP the Hamiltonian (\ref{Hmeanf}) takes  a constant value over all the states
compatible with a given value of the conserved quantity $\sum _i S_i$. Hence all configurations have
equal statistical weight, as in an equilibrium state at infinite temperature, and 
the value of the average domains' size can be trivially computed as  
$\langle \ell \rangle=\sum _{\ell=1}^\infty \ell p(\ell)=2$,
where $p(\ell)=2^{-\ell}$ is the probability to find $\ell$ aligned spins. This explains the initial 
{\it plateau} regime with $L(t)\simeq 2$ observed in Fig. \ref{lt_COP_beta2.5}. It must be noticed that the quantity $\sum_i S_i$ is exactly 
conserved over the whole system, but not on a region of size of order $R$ where the mean field 
solution applies. This explains why on longer times the plateau ends and $L(t)$ keeps growing
in an exponential way, basically for the same reason as for NCOP.

\section{Higher dimensionality}

In this Section we first argue that the processes
inducing a ballistic regime in the non conserved case and a logarithmically slow regime in the conserved case
are still present in higher dimension. 

Let us start with the NCOP model.
As discussed in Ref.~\cite{CLZ}, the asymptotic behaviour $L(t)\sim t^{1/2}$ 
can be understood using the same line of reasoning as for $d=1$. In this case one
considers square or cubic domains of size $L$ 
and asks what is the typical time $t(L)$ to close it.
With nearest neighbor interactions it is shown that $t(L)\sim L^2$ in any dimension.

With the interaction (\ref{coupling}), a spin close to the
domain interface feels an effective field which is always parallel to the majority phase,
much in a way similar to what happens in $d=1$ and discussed in Fig.~\ref{fig1}(a).
Because of this mechanism all spins at the interface are drifted in the direction
to close the domain: since the distance to close it is $L$, if the drift is constant 
$t(L)\sim L$, leading to the ballistic regime. 
As shown by our numerical simulations (see below) the drift decreases with $L$, leading 
to a crossover from ballistic to diffusive regime, as in $d=1$.  

\begin{figure}[h]
\begin{center}
\includegraphics*[width=0.95\columnwidth]{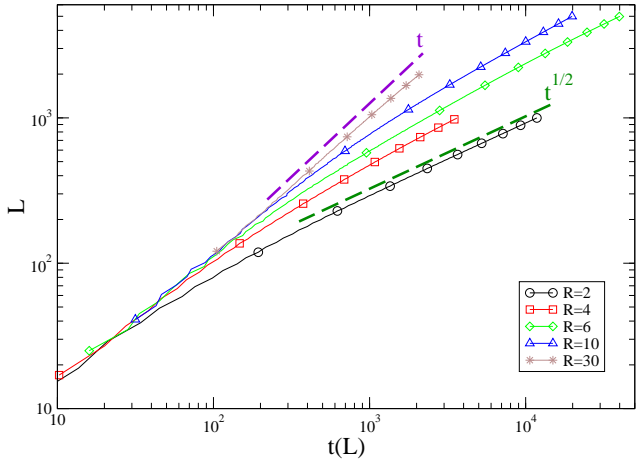}
\end{center}
\caption{The closure time $t$ (horizontal axis) of a square bubble of linear size $L$ 
(vertical axis) is plotted on a double logarithmic scale for a two-dimensional system at $T=0.5$. 
Different curves correspond to the 
  coupling (\ref{coupling}) with different values of the interaction range $R$ (see the legend).
  The dashed green and violet lines are the algebraic forms $t^{1/2}$ and $t$,
respectively.} 
\label{length_D2_singl}
\end{figure}

We have computed numerically the closure time $t(L)$
needed to reverse all spins in a square droplet. 
In Fig. \ref{length_D2_singl} one clearly sees that, while for small values of $R$ the asymptotic
regime $L(t)\propto t^{1/2}$ is  quickly entered, for larger values of $R$ a crossover is observed
between an early ballistic regime with $L(t)\sim t$ and the asymptotic diffusive one. 
Notice that,
at variance with the $d=1$ case, there is no indication of the {\it logarithmic} regime after the ballistic one.
This can presumably be ascribed to the difficulty (or impossibility) to arrange the model parameters 
as to open the time window where such regime would live.

\begin{figure}[h]
\begin{center}
\includegraphics*[width=0.95\columnwidth]{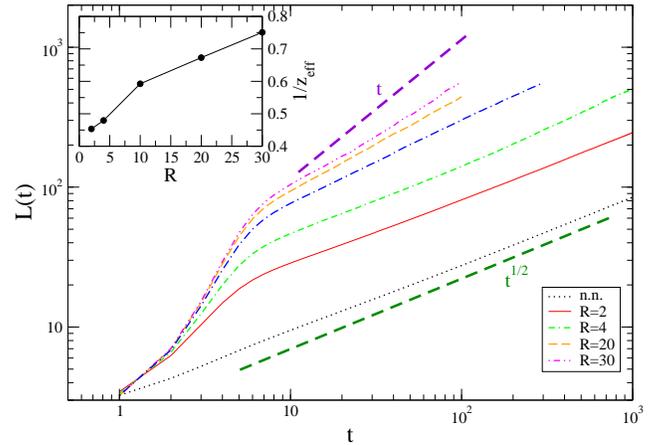}
\end{center}
\caption{$L(t)$ is plotted against time on a double logarithmic
  scale for a quench with NCOP to $T_f=0.5$ in $d=2$.
  Different curves correspond to the next neighbors (n.n.) interaction and to the 
  coupling (\ref{coupling}) with different values of the interaction range $R$ (see the legend).
  The dashed green and violet lines are the algebraic forms $t^{1/2}$ and $t$,
respectively.
Inset: Evaluation of the effective coarsening exponent (see the main text).
}
\label{fig_2d}
\end{figure}

We have then considered the evolution of a two dimensional system with many interfaces
after a quench from $T_i=\infty$: results are shown in Fig. \ref{fig_2d}. 
At short time an exponential increase 
is observed for any value of $R$. This regime, which can be interpreted along the lines of the
$d=1$ case as a mean field effect, extends to larger and larger values of $L(t)$ upon increasing
$R$, as in $d=1$. As compared to the one dimensional case, here such behavior extends 
to somewhat larger values of $L(t)/R$, as expected since the mean field character is enhanced
upon raising $d$. This fact makes the observation of the following preasymptotic regimes very difficult, 
since they are compressed between the long-lasting exponential stage and the 
asymptotic one which sets in shortly after. Our study in $d=1$ and the results in Fig. \ref{length_D2_singl}
suggest that 
the intermediate regimes should be
more easily observed increasing $R$, but this cannot be done at will in simulations.
Despite all the above, one clearly sees that, upon raising $R$, a second regime sets in
where $L(t)$ grows slower than exponentially but faster that $t^{1/2}$. We evaluate the effective exponent
in this stage by fitting $L(t)$ with the power law $t^{1/z_{eff}}$ in the range $t\in [10,100]$.
This quantity is shown in the inset of Fig. \ref{fig_2d}.
The data are compatible with a convergence to $1/z_{eff}=1$ for large $R$. 
This provides evidence
for the existence of a ballistic regime at large $R$, already observed for simple 
configurations (Fig.~\ref{length_D2_singl}), also in the complete system.
In conclusion, data support the expectation that
the different regimes observed in $d=1$, particularly the mean field and the ballistic ones, are present also in $d>1$ for NCOP.

As for the conserved model, numerical simulations in the presence of long range interactions are much more
demanding in $d>1$, but it is still possible to argue that a slowdown of dynamics must appear in the pre-asymptotic regime.
In fact, in the presence of the coupling (\ref{coupling}), whatever is $d$ 
the energy $E(\vec x)$ of a monomer has a maximum somewhere halfway between each pair of clusters, 
see $E(x)$ for $d=1$ above Eq.~(7).  
This maximum represents an energy 
barrier to the diffusion process which hinders the exchange of matter among clusters. 
Therefore, the probability that a monomer 
travels to a larger cluster
(therefore promoting coarsening) vanishes exponentially with the size of the clusters, 
which is also their typical distance. 
In conclusion, we expect a strong slowdown of COP dynamics in $d>1$ as well.

\section{Final considerations}
In this paper we have suggested to interpret the coarsening law $L(t)$ through the
Eq.~(\ref{etL}), where dynamics is characterized by a generalized
drift coefficient $\mu(L)$ and by a generalized diffusion coefficient
${\cal D}(L)$, both dependent on the size $L$ of domains.
The universal coarsening exponents for the NCOP and COP models, $1/2$ and $1/3$ 
respectively, derive from a negligible drift and a diffusion which is constant
for NCOP and it is inversely proportional to $L$ for COP.

The switching of the coupling (\ref{coupling}) allows one to have a model
where the drift is not negligible and the diffusivity has non standard behavior.
In particular, we stress that enlarging the range $R$ of coupling has different effects
according to the absence or to the presence of the conservation law.
In the former case (NCOP), interactions make domain wall
diffusion anisotropic, with a drift favouring the closure of a domain:
this process speeds up coarsening.
In the latter case (COP), interactions make difficult the exchange 
of monomers between neighbouring domains, reducing the diffusivity and slowing
down coarsening.

It is also interesting that both models, NCOP and COP, 
have a logarithmic coarsening regime, see Table~1. 
Besides having a much larger extension for COP, such a regime
has a remarkably different physical interpretation in the two cases. 
In NCOP it is due to an exponentially
small drift and a negligible diffusivity, while in COP it is the other way round,
see Table~1.
This case shows the importance to focus on $\mu(L)$ and ${\cal D}(L)$
in order to understand the dynamics,
an approach that could help elucidating the origin of preasymptotic 
regimes observed in other phase-ordering systems, such as binary alloys~\cite{CC01}, 
systems with quenched disorder~\cite{disorder}, hydrodynamics interactions~\cite{quench},
etc.

Our results open a number of avenues for future investigations in models with a space-decaying
coupling constant, e.g. the analysis of preasymptotic scenarios with an 
interaction decaying algebraically~\cite{algebraic}.
The issue of the modifications to the present picture due to quenched disorder
\cite{disorder}, is another interesting topic worth of investigation.
In addition, to the best of our knowledge the effect of space-decaying interactions on 
the percolation properties of the growing
structure \cite{perc}, on the aging properties \cite{aging},
and on other properties of coarsening systems have never been studied before.

{\bf Acknowledgments}

We thank M. Marsili for useful discussions.
F.C. acknowledges financial support by MIUR PRIN 2015K7KK8L.


\begin{thebibliography}{99}

\bibitem{crp}
F. Corberi and P. Politi, Comptes Rendus Physique, {\bf 16}, Issue 3, 255 (2015).
See also the other contributions in the same issue.

\bibitem{quench}
A.J. Bray, Adv. Phys. {\bf 43} 357 (1994).
A. Onuki, {\it Phase Transition Dynamics}, Cambridge University Press, Cambridge (2004).
S. Puri and V. Wadhawan eds., {\it Kinetics of Phase Transitions}, Taylor and Francis, London (2009).

\bibitem{pattern}
M. Cross and H. Greenside, 
{\it Pattern formation and dynamics in nonequilibrium systems}. (Cambridge University Press, 2009).
A.A. Nepomnyashchy, Comptes Rendus Physique, {\bf 16}, Issue 3, 267 (2015).

\bibitem{condensation}
C. Godr\`eche, in M. Henkel, M. Pleimling, and R. Sanctuary (Eds.), {\it Ageing and the Glass Transition}, 
Lect. Notes Phys. {\bf 716}, Springer (2007). C. Godr\`eche and J-M. Luck, J. Phys.: Condens. Matter 
{\bf 14}, 1601 (2002); Eur. Phys. J. B {\bf 23}(4), 473 (2001).
C. Castellano, F. Corberi, and M. Zannetti, Phys. Rev. E {\bf 56}, 4973 (1997). 
F. Corberi, G. Gonnella, and A. Mossa, Chaos, Solitons and Fractals {\bf 81} 510 (2015).
F. Corberi, Phys. Rev. E {\bf 95}, 032136 (2017).
M. R. Evans, Braz. J. Phys. {\bf 30}, 4257 (2000).

\bibitem{dnls} 

B. Rumpf and A. C. Newell, Phys. Ref. Lett. {\bf 87}, 054102 (2001);
B. Rumpf, Phys. Rev. E {\bf 69}, 016618 (2004);
S. Iubini, A. Politi, P. Politi, J. Stat. Mech. 073201 (2017). 


\bibitem{redner} 
S. Redner, {\it A Guide to First-Passage Processes} (Cambridge University Press, 2001).

\bibitem{exp_sys}

For an experimental example, see L. Liu and G. Bastard,
Phys. Rev. B {\bf 25}, 487 (1982).

\bibitem{CLZ}
F. Corberi, E. Lippiello, and M. Zannetti,
Phys. Rev. E {\bf 78}, 011109 (2008).

\bibitem{algebraic}
A.J. Bray, Phys. Rev. E {\bf 47}, 3191 (1993). A.J. Bray and A.D. Rutenberg, Phys. Rev. E {\bf 49},
R27 (1994). A.D. Ruthenberg and A.J. Bray, Phys. Rev. E {\bf 50}, 1900 (1994).
B.P. Lee and J.L. Cardy, Phys. Rev. E {\bf 48}, 2452 (1993).

\bibitem{CC01}
C. Castellano and F. Corberi, Phys. Rev. B {\bf 63}, 060102 (2001).

\bibitem{disorder}
F. Corberi, Comptes Rendus Physique, {\bf 16}, Issue 3, 332 (2015).
S. Puri, Phase Transitions {\bf 77}, 469 (2004).

\bibitem{perc}
J.J. Arenzon, A.J. Bray, L.F. Cugliandolo, and A. Sicilia, Phys. Rev. Lett. {\bf 98}, 145701 (2007).
A. Sicilia, J.J. Arenzon, A.J. Bray, and L.F. Cugliandolo, Phys. Rev. E {\bf 76}, 061116 (2007).
A. Sicilia, Y. Sarrazin, J.J. Arenzon, A.J. Bray, and L.F. Cugliandolo, Phys. Rev. E {\bf 80}, 031121 (2009).
K. Barros, P.L. Krapivsky, and S.Redner, Phys. Rev. E {\bf 80}, 040101 (2009).
J. Olejarz, P.L. Krapivsky, and S. Redner, Phys. Rev. Lett. {\bf 109}, 195702 (2012).
T. Blanchard and M. Picco, Phys. Rev. E {\bf 88}, 032131 (2013).
T. Blanchard, F. Corberi, L. F. Cugliandolo, and M. Picco, Europhys. Lett. {\bf 106}, 66001 (2014).
F. Corberi, L.F. Cugliandolo, F. Insalata, and M. Picco, Phys. Rev. E {\bf 95}, 022101 (2017).

\bibitem{aging}
J.P. Bouchaud, L.F. Cugliandolo, J. Kurchan and M. Mezard in {\it Spin Glasses and Random fields}, Directions in Condensed Matter Physics {\bf 12}, 161, A.P. Young (Ed.), World Scientific, Singapore, (1998). 
F. Corberi, L.F. Cugliandolo, H. Yoshino, in {\it Dynamical heterogeneities in glasses, colloids, and granular media}, Eds.: L. Berthier, G. Biroli, J-P Bouchaud, L. Cipelletti and W. van Saarloos, Oxford University Press, Oxford (2011).

\end{thebibliography}
\end{document}